\documentclass[11pt, reqno]{amsart}
\usepackage{amsmath,amssymb, times}
\usepackage[all]{xy}
\usepackage[margin=1.1in]{geometry}
\usepackage{pdfpages}
\usepackage{cancel}
\usepackage{svg}
\usepackage{mathtools}
\def\sideremark#1{\ifvmode\leavevmode\fi\vadjust{\vbox to0pt{\vss
\hbox to 0pt{\hskip\hsize\hskip1em%                          
\vbox{\hsize2cm\tiny\raggedright\pretolerance10000%        
\noindent {\color{red}{#1}}\hfill}\hss}\vbox to8pt{\vfil}\vss}}}%

\usepackage{caption}
\usepackage{subcaption}
\theoremstyle{plain}
\newtheorem{propn}{Proposition}[section]
\newtheorem{thm}[propn]{Theorem}
\newtheorem{lemma}[propn]{Lemma}

\newtheorem*{mt1}{Theorem A}
\newtheorem*{mt2}{Theorem B}

\theoremstyle{definition}

\theoremstyle{remark}
\newtheorem*{rem}{Remark}

\usepackage{listings}
\usepackage{xcolor}

\definecolor{codegreen}{rgb}{0,0.6,0}
\definecolor{codegray}{rgb}{0.5,0.5,0.5}
\definecolor{codepurple}{rgb}{0.58,0,0.82}
\definecolor{backcolour}{rgb}{0.95,0.95,0.92}

\lstdefinestyle{mystyle}{
    commentstyle=\color{codegreen},
    keywordstyle=\color{magenta},
    numberstyle=\tiny\color{codegray},
    stringstyle=\color{codepurple},
    basicstyle=\fontfamily{pcr}\selectfont,
    breakatwhitespace=false,         
    breaklines=true,                 
    captionpos=b,                    
    keepspaces=true,                 
    numbers=left,                    
    numbersep=5pt,                  
    showspaces=false,                
    showstringspaces=false,
    showtabs=false,                  
    tabsize=2
}
\usepackage{hyperref} % Used to hyperlink url and email addresses
\hypersetup{pdfborder={0 0 0}, colorlinks=true, urlcolor=blue}

\usepackage{tikz}

\numberwithin{equation}{section}

\begin{document}

\title{Computing $\sqrt{2}$ with FRACTRAN}
\author{Khushi Kaushik}
\address{ Department of Computer Science, California State University Fullerton.}
\author{Tommy Murphy}
\address{Department of Mathematics, California State University Fullerton.}
\email{tmurphy@fullerton.edu}
\urladdr{\href{http://www.fullerton.edu/math/faculty/tmurphy/}{http://www.fullerton.edu/math/faculty/tmurphy/}}
\author{David Weed}
\address{ Department of Mathematics, UC Riverside.}

\begin{abstract}
The FRACTRAN programs $\sqrt{2}$GAME and NR$\sqrt{2}$GAME are presented, both of which compute the decimal expansion of $\sqrt{2}$. Our $\sqrt{2}$GAME is analogous to Conway's  PIGAME program. In fact, our proof carries over to PIGAME to produce a simpler proof of Conway's theorem as well as highlight how the efficiency of the program can be improved. NR$\sqrt{2}$GAME encodes the canonical example of the Newton--Raphson method in FRACTRAN.
 \end{abstract}

\maketitle

\section{Introduction}

Fractran is a Turing complete esoteric programming language with several notable features. It is simple to understand how the language works.  One can code any standard mathematical algorithm in FRACTRAN, and moreover the G\"odel number of any program is straightforward to explicitly compute. Conway developed it \cite{conway} and produced several novel examples, for instance producing PRIMEGAME, which computes, in order, every prime number as well as PIGAME, which generates, in order, the decimal expansion of $\pi$.  PIGAME uses Wallis's infinite product formula for $\frac{\pi}{2}$. As Conway himself states, the proof is nontrivial. It involves using some heavy machinery (e.g. M\"ahler's famed irrationality measure for $\pi$)  to ensure truncating this infinite product after a certain even number $E\geq 4\times 2^{10^n}$ terms is sufficiently accurate to compute the $n$-th digit of $\pi$.

Our first main theorem computes, in order, the decimal expansion of $\sqrt{2}$ via Catalan's \cite{cat} infinite product expansion of $\sqrt{2}$. The mechanics of proof are largely analogous to Conway's, however, we find a simpler proof that our truncated approximation is sufficiently accurate to compute. This simpler proof also carries over to PIGAME:  one then sees \textit{a posteori} that a simpler and faster program could be written to compute the decimal expansion of $\pi$.

There are of course several extremely efficient ways to compute the decimal expansion of $\sqrt{2}$. The most standard is the Newton-Raphson method. Our second theorem presents NR$\sqrt{2}$GAME, which computes the $n$-th digit of $\sqrt{2}$ via this standard algorithm.

\subsection{Acknowledgments} K.K acknowledges support from the LSAMP grant. K.K. and D.W. were supported with a summer undergraduate research grant in 2023 by the Department of  Mathematics at CSU Fullerton. K.K. also thanks the  Undergraduate Research Opportunity Center, CSU Fullerton, for travel support.

\section{Rules of the game}

\subsection{Initial Comments}
Turing Machines model the action of a computer. The main parts of a Turing machine are a way to store data, originally abstractly)thought of as an infinitely long tape, and a set of rules that allow the conditional change of that data. The starting state of the tape is thought of as the input of the code, and the resulting state is the output. Different programs are then made by changing the rules.  In FRACTRAN,  the entirety of the data is stored in a finite set of fractions (the program) and a single integer (the register).   Via the Fundamental Theorem of Arithmetic, each integer 
   $n\in \mathbb{N}$ admits a unique prime decomposition $n = \prod\limits_{i=1}^k p_i^{\alpha_i}$.  Conway's simple idea is to encode a Turing machine starting with a number $N$ so that each power in its prime decomposition gives the state the Turing machine is in. Each power of the prime appearing in $N$ tells us the initial state of our system: it tells us what is in each register.   Now multiply N by a fraction $f_i$ so that $f_iN$ is also a whole number: if we take the prime factor decomposition of the numerator and denominator of  $f_i$,  we have that $f_iN\in \mathbb{N}$ if, and only if, the powers appearing in the prime decomposition of $N$  have been redistributed. Included in this statement is the possibility that the primes with 0 in their register will initially become non-empty.   This exactly models the mechanisms of a Turing machine and is the basic idea underlying FRACTRAN.  More formally, we need an initial state (a stored number) $N\in \mathbb{N}$ which is in our \textit{register} and a fixed list of fractions 
$\lbrace f_1, f_2, \ldots, f_n.\rbrace$. Compute $f_iN$, with $i=1,2, \ldots, n$, until we reach the first instance where  $f_jN\in \mathbb{N}$, $j\in \lbrace 1, 2, \ldots, n\rbrace$. Then change the register to $f_jN$ and iterate. In practice, Conway thinks of FRACTRAN as a flowchart that proceeds from one node (state) to another. To indicate where to go, the nodes are connected by arrows with a well-defined notational hierarchy  as follows:
\begin{center}
\begin{figure}[h]
    \includegraphics[width=.5\textwidth]{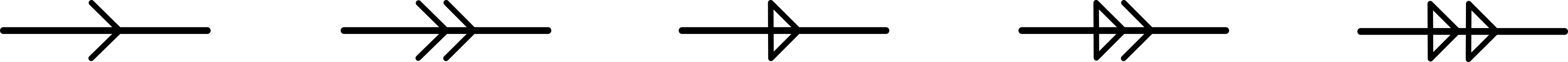}
    \caption{Hierarchy of arrows}
    \end{figure}
\end{center}
These arrows are then labeled with fractions, which tell us how to multiply our register number. There is a well-understood algorithm to convert this flowchart into a list of fractions. We refer the reader to \cite{conway} for full details and to the appendix, where a Python code for this algorithm is presented.  Let us give a simple example and write a program to add $a$ and $b$. Store the numbers via $N=2^a3^b$. Then build a single loop labeled with $2/3$.
\begin{center}
\begin{figure}[h]
\includegraphics[width=.3\textwidth]{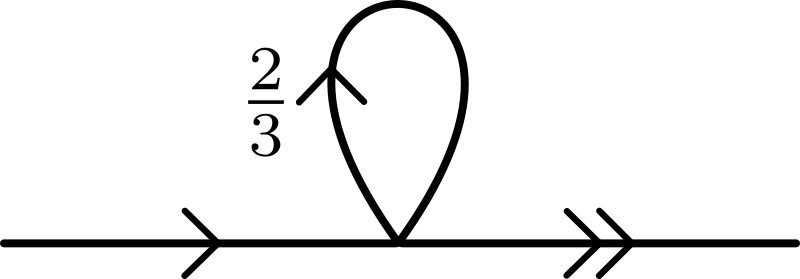}
\caption{A flow-chart for addition}
\end{figure}
\end{center}

 Every time we go around the loop, we multiply $N$ by $\frac{2}{3}$. It is easy to see this game ends with output $2^{a+b}$.  The FRACTRAN code is easy to derive in this example: $N=2^a3^b$ is our initial state, and $\lbrace \frac{2}{3}\rbrace$ is our list of fractions.

\section{Conway's PIGAME}

\begin{thm}(PIGAME \cite{conway})
When started at $2^n\cdot89$, the FRACTRAN code 
\[ \frac{365}{46}\ \frac{29}{161}\ \frac{79}{575}\ \frac{679}{451}\ \frac{3159}{413}\ \frac{83}{407}\ \frac{473}{371}\ \frac{638}{355}\ \frac{434}{335}\ \frac{89}{235}\ \frac{17}{209}\ \frac{79}{122}\ \frac{31}{183} \frac{41}{115}\ \frac{517}{89}\ \frac{111}{83}\ \frac{305}{79}\ \frac{23}{73}\ \frac{73}{71}\]\[\frac{61}{67}\ \frac{37}{61}\ \frac{19}{59}\ \frac{89}{57}\ \frac{41}{53}\ \frac{833}{47}\ \frac{53}{43}\ \frac{86}{41}\ \frac{13}{38}\ \frac{23}{37}\ \frac{67}{31}\ \frac{71}{29}\ \frac{83}{19}\ \frac{475}{17}\ \frac{59}{13}\ \frac{41}{291}\ \frac{1}{7}\ \frac{1}{11}\ \frac{1}{1024}\ \frac{1}{97}\]
will terminate at $2^{\pi(n)}$, where $\pi(n)$ is the $n$-th  digit in the decimal expansion of $\pi$. 
\end{thm}

This list of fractions is  generated via Conway's algorithm from the following  flowchart:

\begin{center}
\begin{figure}[h]
\includegraphics[width=.8\textwidth]{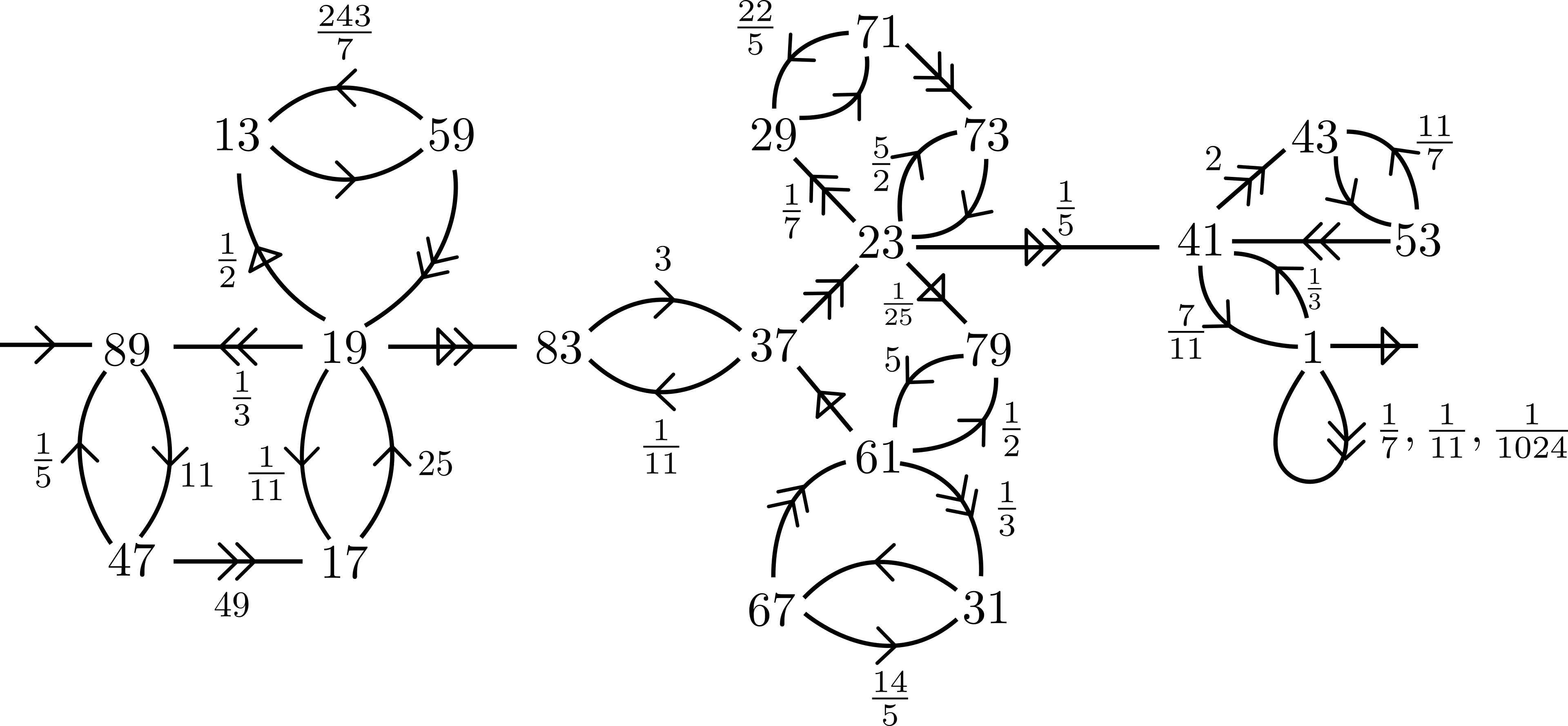}
\caption{The flowchart for PIGAME}
\end{figure}
\end{center}

\begin{rem}
Conway has a convention of arranging the fractions in his code in order of decreasing denominators. There have to be some exceptions to this depending on the code. As explained in the appendix, one can tell which part of the flowchart each fraction corresponds to by looking at the prime decomposition of the numerator and denominator.  For instance, the fraction 79/575 corresponds to moving from node 23 to node 41. However, two arrows are higher in the hierarchy emanating from this node (one going to node 73 and one to node 29), so the fractions corresponding to these arrows must come before 79/575 in the code.
\end{rem}

\begin{rem} There is a small bug in Conway's code, known to experts, where he incorrectly states the code starts at $2^n$. A corrected statement is presented here. 
\end{rem}
Since the proof of Theorem A is based on  PIGAME, and his proof that the algorithm actually works is short on details, it is natural first to discuss the proof and fill in some of the steps. For $n \in \mathbb{N}$, the claim is that running PIGAME will compute the $n$-th decimal digit of $\pi$. The flowchart breaks into three phases. 
\bigskip

{\textbf{Phase 1}} From node 89 to node 83,  the program computes $E$, an even  number $\geq 4\times 2^{10^{n}}$. 

\bigskip

{\textbf{Phase 2}} From node  83 to node 41, the program computes 
\begin{align*}
10^nN_E&= 10^n.2.E(E-2)^2\ldots 4^22^2, \ \text{and} \\
D_E&= (E-1)^2(E-3)^2\ldots 3^21^2.
\end{align*}

{\textbf{Phase 3}} The program computes the integer part of 
$ \frac{10^nN_E}{D_E} $ and reduces it modulo $10$.
\bigskip

The mechanism of these phases are all fully explained in \cite{conway}.  The number computed in Phase 3 is the $n$-th term in the decimal expansion of 
\[\pi_E = \frac{N_E}{D_E} =  \frac{2.E(E-2)^2\ldots 4^22^2}{(E-1)^2(E-3)^2\ldots 3^21^2}.\]  Multiplying the numerator of $\pi_E$ by $10^n$ shifts the decimal unit of $\pi_E$ exactly $n$ places to the right. Taking the floor function turns this into an integer, and reducing mod 10 allows us to find the $n$-th term in the decimal expansion of $\pi_E$. To complete the proof, one has to compute explicitly how close $\pi_E$ is to $\pi$. Another issue to bear in mind comes from the well-known fact that two numbers can be very close together but have differing decimal expansions due to the identification $1 = 0.99\dot{9}$.

So, to show the program actually works, it remains to prove that the   $n$-th decimal digit of $\pi$ and $\pi_E$ agree. To this end, Conway states without proof that 
\begin{equation}\label{e1}
\|\pi - \pi_E\| < \frac{\pi}{E}
\end{equation}
Then $\|\pi - \pi_E\| < \frac{\pi}{E} < 10^{-n}$,  meaning $\pi$ and $\pi_E$  agree to $n$ decimal places unless one of them has a decimal expansion containing only zeros from the $n$-th decimal place onwards (where we make the usual identification $1 = 0.99\dot{9}$). The proof thus reduces to two key steps; (i)  establish Equation (\ref{e1}), and (ii) show that $10^n\pi_E$ cannot be an integer.

\section{ Establishing Equation (\ref{e1})}

The first step is to show that $\pi< \pi_E$ holds for all $E$ even. By way of contradiction, if $\pi_{E_0}<\pi$ then $\pi_{E_0+2}< \pi_{E_0}$, since cancelling common terms we have
\[
\pi_{{E_0+2}}< \pi_{E_0} \iff \frac{E_0(E_0+2)}{(E_0+1)^2} <1
\]
which is true for all $E_0$. Iterating this argument we obtain (with $E =2j$ denoting the subsequence of even integers)
\[
\pi = \lim_{E\rightarrow \infty} \pi_E < \pi_{E_0} < \pi\]
a contradiction. Now for $E$ even, we define
$$
\pi_{\tilde{E}} = \pi_E\left(\frac{E}{E+1}\right)$$
 A directly analogous argument left to the reader shows that $\pi_{\tilde{E}} <\pi$.  Putting these two facts together we obtain
\begin{equation}\label{e2}
\pi_{\tilde{E}} < \pi < \pi_E
\end{equation}

 Equation (\ref{e2}) implies the desired Equation (\ref{e1}). This is a simple computation:
 \[\| \pi - \pi_E \| \overset{\Delta}{= } \pi_E-\pi < \frac{\pi}{E} \iff \pi_{\tilde{E}} < \pi.\]
Note both inequalities in Equation (\ref{e2}) are used. The fact $\pi < \pi_E$ is used for $\overset{\Delta}{=}$. Then   
\[\pi_E-\pi < \frac{\pi}{E} \iff \pi_E < \pi\left(\frac{E+1}{E}\right)\]
which rearranges to the statement that $\pi_{\tilde{E}} < \pi$, i.e. the other inequality in Equation (\ref{e2}).

Now we know $\pi$ and $\pi_E$ are within $10^{-n}$ of each other, it remains to show their decimal expansion agrees in the $n$-th decimal place. To this end, Conway utilizes the following result. 

\begin{lemma}\label{mahler}(M\"ahler's irrationality measure) If $p/q$ is any rational number with $g.c.d.(p,q) = 1$,  \[ \bigg\|\pi - \frac{p}{q}\bigg\| > \frac{1}{q^{42}}.\]
\end{lemma}
 Write $\pi_E= \frac{p}{q}$, with $g.c.d. (p,q) = 1$. Applying M\"ahler's Lemma
 \[\frac{1}{q^{42}} < \bigg\|\pi - \frac{p}{q}\bigg\| < \frac{\pi}{E} < \frac{1}{10^{42n}},\]
 whence (since $x\rightarrow x^{42}$ is an increasing function)  $q> 10^n$. Assume that  \[10^n\pi_E = \frac{10^nN_E}{D_E} = \frac{10^np}{q}\] is an integer. Since  $q>10^n$, there is a prime number $r$ whose multiplicity in the prime decomposition of $q$ is greater than the multiplicity of $r$ in the prime decomposition of $10^n$ (the power of $r$ could be zero in the prime decomposition of $10^n$).  Hence $r$ divides $p$, which is a contradiction as $p$ and $q$ are coprime. This proves fully that Conway's algorithm works.

There is actually an elementary proof that   $\frac{10^np}{q}$ is not an integer. This will  used in the proof of our first main theorem since there is no irrationality measure for $\sqrt{2}$ (it is algebraic).  Supposing $\frac{10^np}{q}$ is an integer means that $q$ divides $10^n$. Since $q$ is odd, that implies $q= 5^j$ where $j \leq n$. However, when we cancel all the common factors in $N_E$ and $D_E$ to get $p$ and $q$, we cannot cancel the largest prime in $D_E$. This is a consequence of   Dirichlet's theorem, which states there must be at least one (odd) prime between $E$ and $E/2$: this number is greater than 5, and no number in $N_E$ divides into it. This is the desired  contradiction

\begin{rem} It is apparent from our discussion that PIGAME can be simplified: there is no need to generate such a large $E$. The size of $E$ is exploited when using M\"ahler's irrationality measure, but we have seen this is not needed.
\end{rem}

\section{$\sqrt{2}$GAME}

Our first main result is as follows. 
\begin{mt1} When started at $2^n\cdot173$, the Fractran code 

\[
\frac{424,375}{173}, \frac{89}{291}, \frac{13}{194}, \frac{4897}{97}, \frac{1243}{89}, \frac{89}{565}, \frac{4949}{113}, \frac{2425}{101}, \frac{101}{1067}, \frac{501}{83},  \frac{83}{1177}, \frac{103}{107},\frac{365}{206}, \frac{103}{73}, \frac{29}{721}, 
\]
\[ \frac{71}{29}, \frac{638}{355}, \frac{4393}{4189},\frac{191}{71},  \frac{73}{71}, \frac{79}{64375},  \frac{1525}{79}, \frac{79}{122},
\frac{67}{183}, \frac{61}{107}, \frac{2669}{67}, \frac{95761}{785}, \frac{157}{131}, \frac{106,609}{3611}, 
\]
%multplying the new bit in phase 2
\[ \frac{149}{1273}, \frac{385,447}{2533}, \frac{149}{139}, \frac{151}{149}, \frac{2329}{7097}, \frac{151}{137}, \frac{67}{151}, \frac{163}{67}, \frac{1541}{29503}, \frac{335}{7009}, \frac{67}{31}, \frac{938}{1333}, \frac{61}{31},\] 
\[\frac{41}{2575}, \frac{7}{943}, \frac{41}{11}, \frac{254}{41}, \frac{53}{127}, \frac{2921}{371},  \frac{53}{41}, \frac{179}{3}, \frac{179}{7}, \frac{179}{1024}, \frac{179}{13}, \frac{179}{17}, \frac{179}{19}, \frac{179}{23},\frac{179}{47}, \frac{1}{179}\rbrace\]

will terminate at $2^{\sqrt{2}(n)}$, where $\sqrt2(n)$ is the $n$-th  digit in the decimal expansion of $\sqrt{2}$.

\end{mt1}

\begin{center}
\begin{figure}[h]
\includegraphics[width=.8\textwidth]{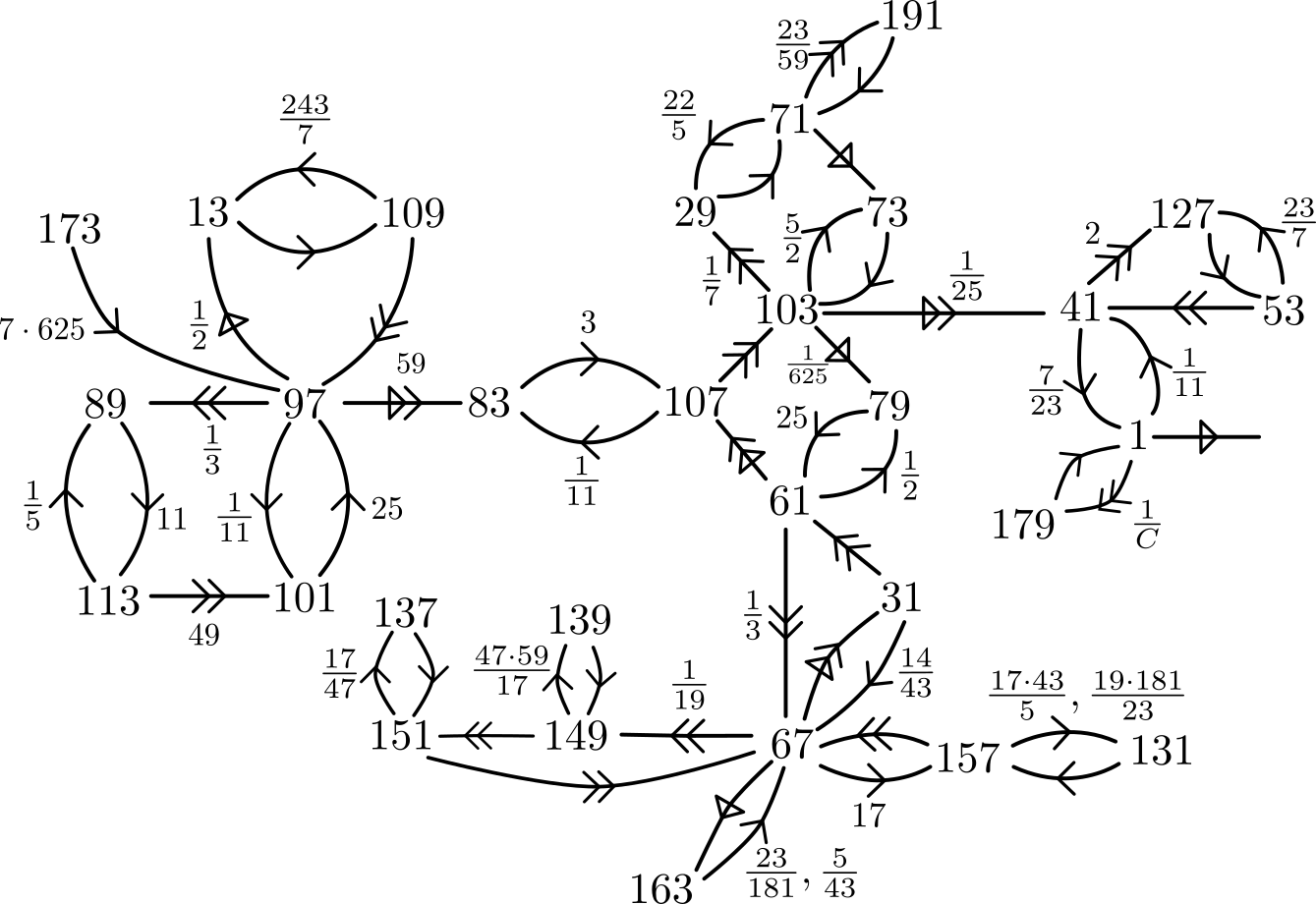}
\caption{The flowchart for $\sqrt{2}$GAME}
\label{figsqrt2game}
\end{figure}
\end{center}

This list of fractions is generated from the flowchart in Figure \ref{figsqrt2game}, where we label each node with a distinct prime number and break all loops up as per Conway's algorithm. For economy of space, $C$ refers to the list of fractions
$$
\frac{1}{3}, \frac{1}{7}, \frac{1}{13}, \frac{1}{17}, \frac{1}{19}, \frac{1}{23}, \frac{1}{47}, \frac{1}{1024}. 
$$

It is obvious from the flowchart that our proof is based on PIGAME. Theorem B, presented in the next section, will show a more standard algorithm for computing $\sqrt{2}$. Nevertheless, $\sqrt{2}$GAME has the merit of fitting into the same framework as PIGAME.

\proof  The starting point is there is an infinite product formula due to Catalan in 1874 \cite{cat} for $\sqrt{2}$ which is very similar to Wallis' formula: viz.
\[
\sqrt{2} = \left(\frac{2.2}{1.3}\right)\left(\frac{6.6}{5.7}\right)\left(\frac{10.10}{9.11}\right)\ldots
\]

This program will truncate Catalan's product formula in an analogous manner to PIGAME. Defining
\[ \sqrt{2}_E: = \frac{2^2}{1.3}\frac{6^2}{5.7}\ldots \frac{(E-4)^2}{(E-5)(E-3)}\frac{E}{E-1} = \frac{N_E}{D_E},\] we claim that the approximation $\sqrt{2}_E$  is sufficiently close to $\sqrt{2}$ and show $10^n\sqrt{2}_E$ is never an integer to prove the program does indeed compute the $n$-th term in the decimal expansion of $\sqrt{2}$.

The first phase, from node 89 to node 83, is almost identical to PIGAME. The only difference is the requirement to kill the extra 2 Conway has in his formula for $N_E$. This comes from the fact that Wallis's infinite product formula is for $\frac{\pi}{2}$, so he needs to double his numerator to approximate $\pi$. In contrast, we have an infinite product formula for $\sqrt{2}$, so a  slight modification of Conway's argument is needed. We skip his first pass around the square region, instead first going into the upper triangular region of Phase 1 with initial values $r_5 = 4$ and $r_7=49$. From here, the first phase  proceeds in the same fashion as Conway, and  it is easy to check  we reach node 83 with 
$$
r_2 = 0, \ r_3 = 1, \ r_5 = E, \ r_7 = 10^n, \ r_{11}=0, \ r_{59}=1
$$
where $E$ is a very large  even number.  For Conway, it suffices to know that each pass around the square region ``at least doubles" $r_5$ (which is initially set to $4$) and keeps it even.  This is because there is a well-defined truncation $\pi_E$ for $\pi$ for any even number $\pi$. However, for Catalan's formula for $\sqrt{2}$ to be appropriately truncated, one has to check that  $E \equiv 2 \ \text{mod}\ 4$. Entering the square region for the first time, we have $r_5=4$, and we exit with $r_4=10$.  In fact if we enter the square region with $r_5 = 4j+2$, we exit with $r_5 = 4(2j+1) + 2$, as the reader can easily check.

In the second phase, from node 83 to node 41, the essential point is that we copy Conway but  modify how much we subtract from $r_5$ during each loop to generate $N_E$. However, we have to also generate $D_E$, which has a different form than the denominator of $\pi_E$.  We break each pass around the region into part (i), where we travel up from node 107, and part (ii), going down from node 61.   Just as in PIGAME, part (i) sets $r_7=0$ and multiples $r_5$ by $r_7$, storing this number in $r_{11}$.  However we also transfer $r_{59}$ to $r_{23}$ and reset $r_{59}=0$. 

Moving onto part (ii) of our loop, the hierarchy of arrows (corresponding to the order we carry out the operations) becomes more delicate. As in PIGAME, we transfer $r_{11}$ to $r_3$, while preserving $r_{11}$. Note   $r_5$ decreases by 2 (as opposed to Conway, who decreases $r_5$ by 1). With this modified value of $r_5$, we multiply $r_3$ by $r_5$, storing this in $r_7$. Then we add $1$ to $r_5$  (storing it in $r_{17}$), and multiply this new number by $r_{23},$ storing it in $r_{59}$.The program continues in this phase until \( r_{5} \) reaches a value of 2. The program then starts phase (i) of the final loop, but cannot go to phase (ii) of the loop and exits to start phase (iii) at node 41.   At the end of the second phase $r_{11} = 10^nN_E$  and $r_{23} = D_E$.  In the following chart, we summarize how each register updates during the second phase, breaking each loop into (i) and (ii) schematically.

\begin{center}
\begin{tabular}[!htb]{|c|c|}
\hline
   \textrm{up (i)} & \textrm{down (ii)} \\ \hline
    $r_{11} = r_7.r_5$ & $r_3 = r_{11}$ \\  \hline
    $r_5 = r_5$ & $r_5 = r_5-2$
    \\  \hline
    $r_7 = 0$ & $r_7 = r_3\cdot r_5$  \\  \hline
    $r_{23} = r_{59}$ & $r_{17} = r_{5}+1$
    \\
     \hline
    $r_{59} = 0$ & $r_{59} = r_{23}\cdot r_{17}$  \\ \hline
    \end{tabular}
\end{center}
\bigskip

To clarify the proof, let us explicitly perform four loops of the second phase (numbered I-- IV) in the following table. Each loop is  broken into parts(i) and (ii). 

\begin{center}
{\tiny{
\begin{tabular}[!htb]{|c|c|c|c|c|c|}
\hline
\textrm{I(i)}&\textrm{I(ii)}&\textrm{II(i)}&\textrm{II(ii)}&\textrm{III(i)}&\textrm{III(ii)}\\
\hline
\hline
$r_{11} = 10^nE$ & $r_5 = E-2$ & $r_{11} = (E-2)^2$ & $r_5 = E-4$ & $r_{11} = 10^n(E)(E-4)^2$ &$r_5 = E-6$  \\
 $r_5 = E$ & $r_7 = E-2$ &$r_5 = E-2$ & $r_7 = 10^n(E)(E-4)$ & $r_5 = E-4$ & $r_7 = (E-2)^2(E-6)$ \\
$r_{23} = 1$ & $r_{17} = E-1$ & $r_7 = 0$ & $r_{17} = E-3$ & $r_7 = 0$ & $r_{17} = E-5$ \\
$r_7 = r_{59} = 0$ & $r_3 = 10^nE$ & $r_{23} = E-1$ & $r_3 = (E-2)^2$ & $r_{23} = (E-1)(E-3)$ & $r_3 = 10^n(E)(E-4)^2$ \\
 & $r_{59} = E-1$  & & $r_{59} = (E-1)(E-3)$ & & $r_{59}= (E-1)(E-3)(E-5)$ \\
 \hline
 \hline
\end{tabular}
}}
\end{center}
\begin{center}
{\tiny
\begin{tabular}[!htb]{|c|c|}
\hline
\textrm{IV(i)}&\textrm{IV(ii)}\\
\hline
\hline
$r_{11} = (E-6)^2(E-2)^2$ & $r_5 = (E-8)$  \\
$r_5=(E-6)$ & $r_7 = 10^n(E)(E-4)^2(E-8)$\\
$r_7 = 0$ & $r_{17} = (E-7)$\\
$r_{23} = (E-1)(E-3)(E-5)$ & $r_{59} = (E-1)(E-3)(E-5)(E-7)$  \\
$r_{59} = 0$ & $r_3 = (E-6)^2(E-2)^2$ \\
\hline
\hline
\end{tabular}
}
\end{center}

Continuing on one  more loop and recording the key register of interest, note that at the end of  loop V(i)  we have $r_{11} = 10^nE(E-4)^2(E-8)^2$. Since $r_5 \equiv 2 \ \text{mod} 4$,   the program will complete an even number   of full loops until $r_5=2$. It will then go into phase (i) of an odd numbered loop, but it cannot go into phase (ii) and so the program passes to  the third phase  with $r_{11} = 10^nN_E$ as claimed.

Moving into the third phase, we copy with obvious modifications the third phase of PIGAME to compute the $n$-th decimal of $\sqrt{2}_E$.  The balance of the proof will involve two steps. 

{\textbf{Step 1}} Establish the inequality 
\begin{equation}\label{e4}
\|\sqrt{2} - \sqrt{2}_E\|< \frac{2\sqrt{2}}{E}.
\end{equation}

{\textbf{Step 2}} Show that  $10^n\sqrt{2}_E$ is not an integer. 
\bigskip

Since $\frac{2\sqrt{2}}{E} < 10^{-n}$, Step 1 and 2  together prove that the program computes the $n$-th term in the decimal expansion of $\sqrt{2}$, just as in PIGAME.  To prove  Equation (\ref{e4}), firstly we show $\sqrt{2}_E> \sqrt{2}$. If to the contrary $\sqrt{2}_{E_0}<\sqrt{2}$ for some $E_0 = 4j+2$, where $j\in \mathbb{N}$, then
\[
\sqrt{2}_{E_0+4} < \sqrt{2}_{E_0} \iff E_0^2 + 4E_0 < E_0^2 + 4E_0 + 3
\]
which is obviously true. Since the sub-sequence $\sqrt{2}_{4j+2} \rightarrow \sqrt{2}$, we again easily derive a contradiction in the exact same manner as in the proof of PIGAME. Setting 
\[\sqrt{2}_{\tilde{E}} = \left(\frac{E}{E+2}\right)\sqrt{2}_E,\]
an analogous proof shows that $\sqrt{2}_{\tilde{E}} < \sqrt{2}$.
In summary for all $E = 4j+2$ we have
\begin{equation}\label{e5}
\sqrt{2}_{\tilde{E}} < \sqrt{2} < \sqrt{2}_E.
\end{equation}
However, Equation (\ref{e5}) is equivalent to Equation (\ref{e4}), as
\[ |\sqrt{2} - \sqrt{2}_E| \overset{\Delta}{= } \sqrt{2}_E - \sqrt{2} < \frac{\sqrt{2}}{x} \iff \sqrt{2}_E < \sqrt{2}\left(\frac{x+1}{x}\right)\]
where $\overset{\Delta}{= }$ uses the second inequality of Equation (\ref{e5}). However Equation (\ref{e5}) establishes $\sqrt{2}_{\tilde{E}} < \sqrt{2}$.  Choose now $x = E/2$ to obtain  
\[\sqrt{2}_{\tilde{E}}= \left(\frac{x}{x+1}\right)\sqrt{2}_E\]
and  we see that Equation (\ref{e4}) immediately follows from the first inequality in Equation (\ref{e5}). 

The final step is to prove $\frac{10^np}{q}$ is never an integer, where $\sqrt{2}_E = \frac{p}{q}$ with $p$ and $q$ coprime. This is directly analogous to our explanation for PIGAME, and the proof of  Theorem A is now complete. \endproof

\section{NRsqrt2 game}
\subsection{Description} The standard way to approximate $\sqrt{2}$ is to use Heron's algorithm, or equivalently the Newton--Raphson method applied to the function $f(x) = x^2-2$ with initial guess $x_1 = p_1/q_1 = 1/1$. This updates via 
\[ x_{k+1} = \frac{p_{k+1}}{q_{k+1}} = \frac{p_k^2 + 2q_k^2}{2p_kq_k}.\]
We claim that computing $x_{2n}$ will generate a rational number sufficiently close to $\sqrt{2}$ to agree to $n$ decimal places. Encoding this as a FRACTRAN program is our second main result. 

\begin{mt2}\label{maint2}
Starting at $2^n\cdot89$, the following FRACTRAN code terminates at $2^{\sqrt{2}(n)}$. 

\[  \lbrace\frac{4979909}{89}, \frac{227,123,851}{466}, \frac{233}{239}, \frac{11809}{23533}, \frac{241}{251}, \frac{60,993}{1687}, \frac{267}{723}, \frac{267}{257}, \frac{17355}{2827},\frac{277}{267},  \frac{271}{277}, \frac{3047}{1355}, \frac{241}{277}, \frac{2959}{1205}, \]
%phase 2 starts after the first two fractions, which are the end of phase 1
\[  \frac{233}{241}, \frac{283}{233}, \frac{859579}{8207},  \frac{283}{281}, \frac{24278273}{18961}, \frac{307}{283}, \frac{313}{5833}, \frac{3170}{2191}, \frac{313}{317}, \frac{331}{313}, \frac{2359}{1655}, \frac{307}{331}, \frac{311}{307}, \frac{8903}{622}, \frac{347}{307}\]
%Phase 3
\[ \frac{359}{14227}, \frac{3350}{15437}, \frac{359}{353}, \frac{367}{359}, \frac{16039}{16515}, \frac{367}{347}, \frac{17101}{694}, \frac{379}{347}, %phase 4  
\frac{397}{9475},  \frac{389}{397}, \frac{3970}{18,283},  \frac{409}{397},  \frac{401}{409}, \frac{19223}{2005},  \frac{379}{409},\]
%The loop and phase 6
\[\frac{421}{379}, \frac{12151}{2947},  \frac{421}{419}, \frac{283}{40837}, \frac{467}{421}, \frac{433}{13543},  \frac{6465}{4763}, \frac{433}{431}, \frac{443}{433}, \frac{5423}{2215}, \frac{443}{439}, \frac{467}{443}
\]
%endgame
\[\frac{457}{467}, \frac{7}{30619}, \frac{457}{3}, \frac{922}{457}, \frac{5093}{3227}, \frac{461}{463}, \frac{457}{463}, \frac{449}{1024}, \frac{449}{3}, \frac{449}{67},  \frac{1}{449}.\rbrace\]

\end{mt2}
\bigskip

\begin{center}
\begin{figure}[h]
\includegraphics[width=.99\textwidth]{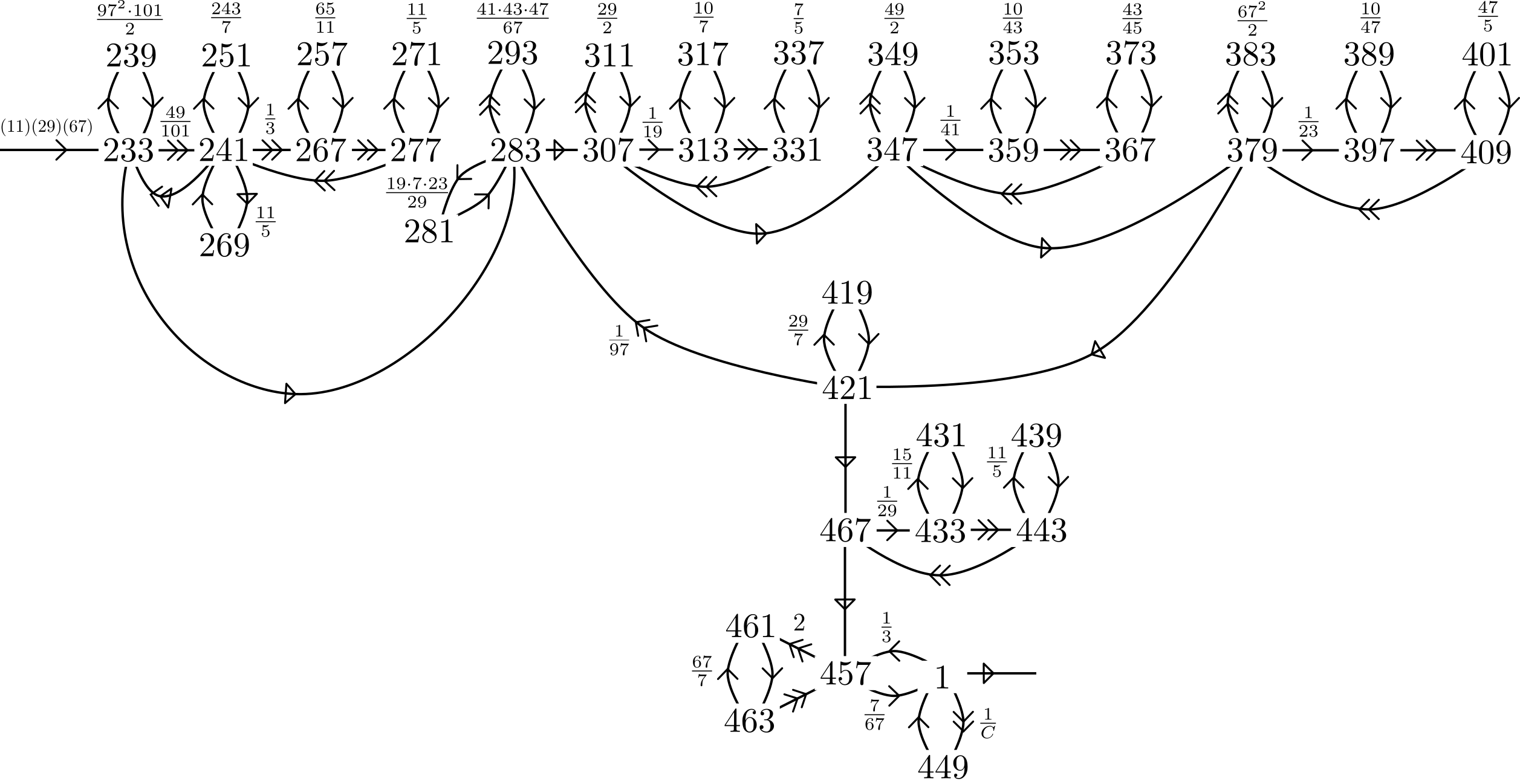}
\caption{The flowchart for NR$\sqrt{2}$GAME}
\label{nrfigure}
\end{figure}
\end{center}

This list of fractions was generated from the flowchart in Figure \ref{nrfigure} following Conway's algorithm.  The program initializes with $r_2=n$. For the reader's ease, we will break the flowchart into phases (i), (ii), etc., in descending order.  Phase (i) terminates at the $283$ node. Initially the algorithm sets $r_{11} = r_{29} = r_{67} = 1$. Our initial guess for $\sqrt{2}$ is $x_1 =r_{29}/r_{67} = 1/1$. In general, we set  $x_j = p_j/q_j$, where  $p_j = r_{29}$ and $q_j = r_{67}$.  As we move along phase 1, note we end with $r_{97} = 2n$. This is the number of iterations of the NR algorithm we must perform. Phase (i) also sets $r_{11} = 10^n$ as the reader can check.

In Phase (ii), we transfer both $p_j$ and $q_j$, where $1\leq j \leq 2n$ is the current stage of the NR algorithm,  to three new registers to facilitate computation later. Phase (ii) ends when we reach the $347$ node.  Here $r_{29} = p_j^2.$ Similarly phase (iii) ends when we reach the $379$ node with $r_7 = 2q^2$, and phase (iv) ends at the $379$ node with $r_{67} = 2p_jq_j$.  We then travel back up to phase (ii) with 
$$
r_{29} = p_j^2 + 2q_j^2 \ \ \ r_{67} = 2p_jq_j \ \ \ \text{and} \ r_5 = 10^n.
$$
This key step encodes the iterative loop. Here $r_{13}$ decreases by 1, and we updated from $x_k$ to $x_{k+1}$ as we travel back through the flowchart to phase (ii).

After $2n$ iterations, we move to phase (v) at node $467$. Here we firstly multiply $r_{29} = p_{2n}$ by $r_{11} = 10^n$ (storing it in $r_3$) before, \'a la PRIMEGAME, computing the integer part of $\frac{10^np_{2n}}{q_{2n}}$ modulo 10.  This is the $n$-th term in the decimal expansion of $x_{2n}$.  Now the mechanisms of the algorithm have been explained, we have to prove that $x_{2n}$ is sufficiently close to $\sqrt{2}$ that both numbers agree in the $n$-th decimal place. To this end, we need the following lemma. 

\begin{lemma}\label{error}
    Suppose $f$ is a smooth function on $[1,2]$ with $\|f'\| \geq L$ and $\|f''\|<M$ for some $L, M>0$. If $f(r) = 0$, then the error  which arises from applying the Newton-Raphson algorithm to f $N$ times starting at $x_1$ is given by \[
    \|x_{N+1} - r \| <\frac{M}{2L}\| x_N - r\|^2. \]
\end{lemma}

\proof (of Theorem \ref{maint2}) To compute the $n$-th decimal digit of $\sqrt{2},$  we need to  estimate  $\epsilon_N := \|x_{N} - \sqrt{2}\|$. For later use  $\epsilon_1 = \frac{1}{2}$. By induction,  it is clear that $x_k \in [1,2]$ for all $k\in \mathbb{N}$. Applying the standard error estimates for Newton's method with $f(x) = x^2-2$, we have $|f''| = 2$ and $2\leq |f'(x)|\leq 4$ on $[1,2]$. We iterate the error bound from Lemma \ref{error} to obtain
\[
\epsilon_{N} < \left(\frac{1}{2}\right)\epsilon_{N-1}^2 < \left(\frac{1}{4}\right)\epsilon_{N-2}^4 \ldots < \frac{\epsilon_1^{2N}}{2^N} = \frac{1}{2^{3N}}.\]
This shows that 
\[
\epsilon_{N} < \frac{1}{10^{n+1}} \iff \frac{1}{2^{3N}} < \frac{1}{10^{n+1}}  
\]
which clearly holds if $N=2n$. With this error bound $x_{N}$ and $\sqrt{2}$ must agree up to the $n$-th decimal place. \endproof

\section*{Appendix: Converting a flowchart into a FRACTRAN code}

The following code converts our flowcharts into their corresponding list of FRACTRAN fractions. For a single node in the flowchart, we write a line describing to what node it is connected and through what fraction. Line 17 then shows how you convert that line into a series of fractions. This is described more in section 8 of \cite{conway}. In a single line the order in which the connections are listed handles the hierarchy of the arrows. 

\begin{lstlisting}[language=Python]
#Each line should be  P, a/b->Q, c/d->R, ...

with open('fracn.txt') as file:
    fractionList = list()
    fractionFactored = list()
    for line in file:
        entries = line.split(', ')
        P = entries[0]
        for entry in entries[1:]:
            print(entry)
            a,temp = entry.split('/')            
            b,Q = temp.split('->')
            Q = Q.strip()
            fractionFactored.append(a+'*'+Q+'/'+b+'*'+P)
            aQ = str(int(a)*int(Q))
            bP = str(int(b)*int(P))
            fractionList.append(aQ+'/'+bP)
    print(fractionFactored)
    print(fractionList)

with open('fractions.txt', 'w') as file:
    file.write(', '.join(fractionList))

with open('fractionsFactored.txt', 'w') as file:
    file.write(', '.join(fractionFactored))
\end{lstlisting}


\newpage



\begin{thebibliography}{99}
\bibitem{cat} Catalan, E. \textit{Sur la constante d'Euler et la fonction de Binet}, C.R. Acad. Sci. Paris S\'er. I Math. 77 (1873), 198-201. 
\bibitem{conway} Conway, J.H. \textit{FRACTRAN: a simple universal programming language for arithmetic}, Open problems in Communication and Computation, Springer-Verlag New York Inc. (1987),  4--26.
\bibitem{sondow} Sondow, J. and Hi, H. \textit{New Wallis- and Catalan-type infinite products for $\pi$, $e$ and $\sqrt{2+\sqrt{2}}$, Amer. Math. Monthly, 117 (2010), no. 10, 912-917.}
\end{thebibliography}
\end{document}